\documentclass[aps,prl,floatfix,twocolumn,superscriptaddress]{revtex4-1}

\usepackage[]{graphicx}
\usepackage{color}

\usepackage{amsmath,amssymb,amsfonts}

\usepackage{slashed}

\begin{document}

\title{Flavor hierarchy of jet quenching in relativistic heavy-ion collisions}

\author{Wen-Jing Xing}
\affiliation{Institute of Particle Physics and Key Laboratory of Quark and Lepton Physics (MOE), Central China Normal University, Wuhan, Hubei, 430079, China}

\author{Shanshan Cao}
\affiliation{Department of Physics and Astronomy, Wayne State University, Detroit, MI 48201, USA}

\author{Guang-You Qin}
\affiliation{Institute of Particle Physics and Key Laboratory of Quark and Lepton Physics (MOE), Central China Normal University, Wuhan, Hubei, 430079, China}

\author{Hongxi Xing}
\affiliation{Institute of Quantum Matter and School of Physics and Telecommunication Engineering,South China Normal University, Guangzhou 510006, China}

\date{\today}
\begin{abstract}

Relativistic heavy-ion experiments have observed similar quenching effects for (prompt) $D$ mesons compared to charged hadrons for transverse momenta larger than 6-8~GeV, which remains a mystery since heavy quarks typically lose less energies in quark-gluon plasma than light quarks and gluons.
Recent measurements of the nuclear modification factors of $B$ mesons and $B$-decayed $D$ mesons by the CMS Collaboration provide a unique opportunity to study the flavor hierarchy of jet quenching. 	
Using a linear Boltzmann transport model combined with hydrodynamics simulation, we study the energy loss and nuclear modification for heavy and light flavor jets in high-energy nuclear collisions.
By consistently taking into account both quark and gluon contributions to light and heavy flavor hadron productions within a next-to-leading order perturbative QCD framework, we obtain, for the first time, a satisfactory description of the experimental data on the nuclear modification factors for charged hadrons, $D$ mesons, $B$ mesons and $B$-decayed $D$ mesons simultaneously over a wide range of transverse momenta (8-300~GeV).
This presents a solid solution to the flavor puzzle of jet quenching and constitutes a significant step towards the precision study of jet-medium interaction.
Our study predicts that at transverse momenta larger than 30-40~GeV, $B$ mesons also exhibit similar suppression effects to charged hadrons and $D$ mesons, which may be tested by future measurements.

\end{abstract}
\maketitle

{\it Introduction} -- Large transverse momentum ($p_{\rm T}$) jets are hard probes of the strongly-coupled quark-gluon plasma (QGP) created in relativistic heavy-ion collisions \cite{Wang:1991xy, Qin:2015srf, Blaizot:2015lma, Majumder:2010qh, Gyulassy:2003mc}.
During their propagation through the QGP medium, jet partons tend to lose energies via elastic and inelastic interactions with the medium constituents, which is usually referred to as jet quenching.
Jet quenching not only leads to the yield suppression for high $p_{\rm T}$ hadrons \cite{Khachatryan:2016odn, Acharya:2018qsh, Aad:2015wga, Burke:2013yra, Buzzatti:2011vt, Chien:2015vja, Andres:2016iys, Cao:2017hhk, Zigic:2018ovr} and full jets \cite{Adam:2015ewa, Aad:2014bxa, Khachatryan:2016jfl, Qin:2010mn, Young:2011qx, Dai:2012am, Wang:2013cia, Blaizot:2013hx, Mehtar-Tani:2014yea, Cao:2017qpx, Kang:2017frl, He:2018xjv}, but also modifies jet-related correlations  \cite{Aad:2010bu, Chatrchyan:2012gt, Qin:2009bk, Chen:2016vem, Chen:2016cof, Chen:2017zte, Luo:2018pto, Zhang:2018urd, Kang:2018wrs} and the internal structures of full jets \cite{Chatrchyan:2013kwa, Aad:2014wha, Chang:2016gjp, Casalderrey-Solana:2016jvj, Tachibana:2017syd, KunnawalkamElayavalli:2017hxo, Brewer:2017fqy, Chien:2016led, Milhano:2017nzm}, as compared to proton-proton collisions.
With the increase of collision energy by more than a factor of 10 from the Relativistic Heavy-Ion Collider (RHIC) to the Large Hadron Collider (LHC), we can now produce abundant jets (and hadrons) with $p_{\rm T}$ of hundreds of GeV, which enables us to peform more and more precise jet quenching studies for heavy-ion collisions.

Heavy (charm and bottom) quarks, due to their finite masses, are expected to lose less energies in QGP than light quarks (and also gluons due to different color factors).
Thus one expects heavy flavor hadrons (e.g., $D$ and $B$ mesons) would exhibit less quenching effects than light charged hadrons.
There has been tremendous effort devoted to heavy quark dynamics in relativistic heavy-ion collisions \cite{Dong:2019byy, Rapp:2018qla, Cao:2018ews, Uphoff:2011ad, He:2011qa, Young:2011ug, Alberico:2011zy, Nahrgang:2013saa, Cao:2013ita, Cao:2015hia, Das:2015ana, Song:2015ykw, Cao:2016gvr, Prado:2016szr, Cao:2017crw, Liu:2017qah, Li:2018izm, Ke:2018tsh, Katz:2019fkc}.
However, experiments have observed similar quenching effects for (prompt) $D$ mesons as compared to charged hadrons at $p_{\rm T}>$~6-8~GeV \cite{Adare:2014rly, Adamczyk:2014uip, ALICE:2012ab}.
Such result challenges our theoretical understanding of the flavor dependence of jet-medium interaction and parton energy loss, and is usually denoted as the flavor hierarchy puzzle of jet quenching.
Reference~\cite{Djordjevic:2013pba} tried to solve this puzzle by suggesting that different patterns in parton fragmentation functions may play important roles in the final-state hadron suppression.
But the final hadron modification pattern also strongly relies on the $p_{\rm T}$ dependence of jet suppression.
Studies in Refs.~\cite{Norrbin:2000zc, Aad:2012ma, Huang:2013vaa, Huang:2015mva, Cao:2015kvb, Kang:2016ofv} indicate that gluons could also contribute to heavy flavor jet and hadron productions.
References~\cite{Cao:2016gvr, Cao:2017hhk} have built a systematic framework to study the evolution of heavy and light jet partons in QGP on the same footing, but neglected the gluon contributions to heavy flavor hadron productions.
Therefore, a satisfactory description of the nuclear modifications of light charged hadrons, $D$ mesons and $B$ mesons together is still lacking.
Recently, CMS Collaborations have measured the nuclear modification factors for both $B$ mesons and $B$-decayed $D$ mesons in Pb+Pb collisions at 5.02A~TeV \cite{Sirunyan:2017oug, Sirunyan:2018ktu}; this provides a remarkable opportunity to pin down the flavor dependence of jet quenching in relativistic nuclear collisions.

The objective of our work is to present a comprehensive study on the nuclear modification of both heavy and light flavor jets in high-energy heavy-ion collisions, and to tackle the flavor hierarchy puzzle of jet quenching.
In this work, a next-to-leading-order (NLO) perturbative QCD framework is used to calculate the productions of high $p_{\rm T}$ jet partons and hadrons, allowing a consistent treatment of quark and gluon fragmentations to light and heavy flavor hadrons.
A linear Boltzmann transport (LBT) model is utilized to describe jet evolution in the QGP medium, including both elastic and inelastic interactions between jet partons and the medium constituents.
A relativistic hydrodynamics model is employed to simulate the dynamical evolution of the QGP fireball.
By combining all important ingredients into our state-of-the-art perturbative QCD based jet quenching model, we obtain the first simultaneous description of the experimental data for the nuclear modifications of charged hadrons, $D$ mesons, $B$ mesons and $B$-decayed $D$ mesons over the widest range of transverse momenta ($p_{\rm T}=$~8-300~GeV) in literature.
Our study shows that, due to the mass effect, $B$ mesons typically exhibit less suppression than charged hadrons and $D$ mesons at not-very-high $p_{\rm T}$. But the mass effect diminishes with increasing $p_{\rm T}$; at $p_{\rm T}>$~30-40~GeV, charged hadrons, $D$ and $B$ mesons all have similar quenching effects. This can be tested by future measurements.


{\it Jet quenching framework} -- We use the NLO framework developed in Refs. \cite{Jager:2002xm, Aversa:1988vb} to calculate jet and high-$p_{\rm T}$ hadron productions in relativistic nuclear collisions. The differential cross section for hadron production in proton-proton collisions can be expressed as follows:
\begin{eqnarray}
d\sigma_{pp\to hX} &=& \sum_{abc} \int dx_a \int dx_b \int dz_c
f_{a}(x_a) f_{b}(x_b)   \nonumber\\
&& \times d\hat{\sigma}_{ab\to c} D_{h/c}(z_c).
\end{eqnarray}
In the above equation, $\sum_{abc}$ sums over all parton flavors, $f_{a}(x_a)$ and $f_{b}(x_b)$ denote parton distribution functions (PDFs) for two incoming partons, $d\hat{\sigma}_{ab\to c}$ is the NLO partonic scattering cross section, and $D_{h/c}(z_c)$ represents the parton-to-hadron fragmentation function (FF).
The PDFs are taken from CTEQ parameterizations \cite{Pumplin:2002vw}, and the FFs are taken from Ref. \cite{Kretzer:2000yf} for charged hadrons, Ref. \cite{Kneesch:2007ey} for $D$ mesons, and Ref. \cite{Kniehl:2008zza} for $B$ mesons.
Note that in the NLO framework \cite{Jager:2002xm, Aversa:1988vb}, one has to include both quark and gluon fragmentions to heavy and light hadron productions at high $p_{\rm T}$.

For jet and high-$p_{\rm T}$ hadron productions in relativistic heavy-ion collisions, we need to account for two different nuclear effects. The initial-state nuclear shadowing effect is taken into account by applying EPS09 parameterizations \cite{Eskola:2009uj} for nuclear PDFs.
The final-state hot medium effect is the focus of our work: high-energy jet partons experience elastic and inelastic interactions with the strongly-coupled QGP before fragmenting into high $p_{\rm T}$ hadrons.
The hot medium effect is incorporated in our study by using the LBT approach developed in Refs. \cite{Cao:2017hhk, Cao:2016gvr, He:2015pra}.

In the LBT model, the evolution of jet partons in the QGP medium is simulated according to the following Boltzmann equation:
\begin{eqnarray}
p_a \cdot \partial f_a(\mathbf{x}, \mathbf{p}_a, t) = E_a(\mathcal{C}_{\rm el} + \mathcal{C}_{\rm inel})
\end{eqnarray}
where $\mathcal{C}_{\rm el}$ and $\mathcal{C}_{\rm inel}$ denote the collision integrals of elastic and inelastic processes experienced by the parton $a$.

For elastic scatterings between jet partons and medium constituents, we take leading-order $2 \to 2$ perturbative QCD matrix elements $|M_{ab\to cd}|^2$ to calculate the elastic scattering rate $\Gamma_{\rm el}^a = \sum_{bcd} \rho_b(\mathbf{x}) \sigma_{ab\to cd}$ and the elastic scattering probability $P_{\rm el}^a = 1- e^{-\Gamma_{\rm el}^a \Delta t}$ for a given time step $\Delta t$, where $\rho_b(\mathbf{x})$ is the parton density in the QGP medium and $\sigma_{ab\to cd}$ is the parton scattering cross section.

\begin{figure}[tb]
\includegraphics[width=0.49\linewidth]{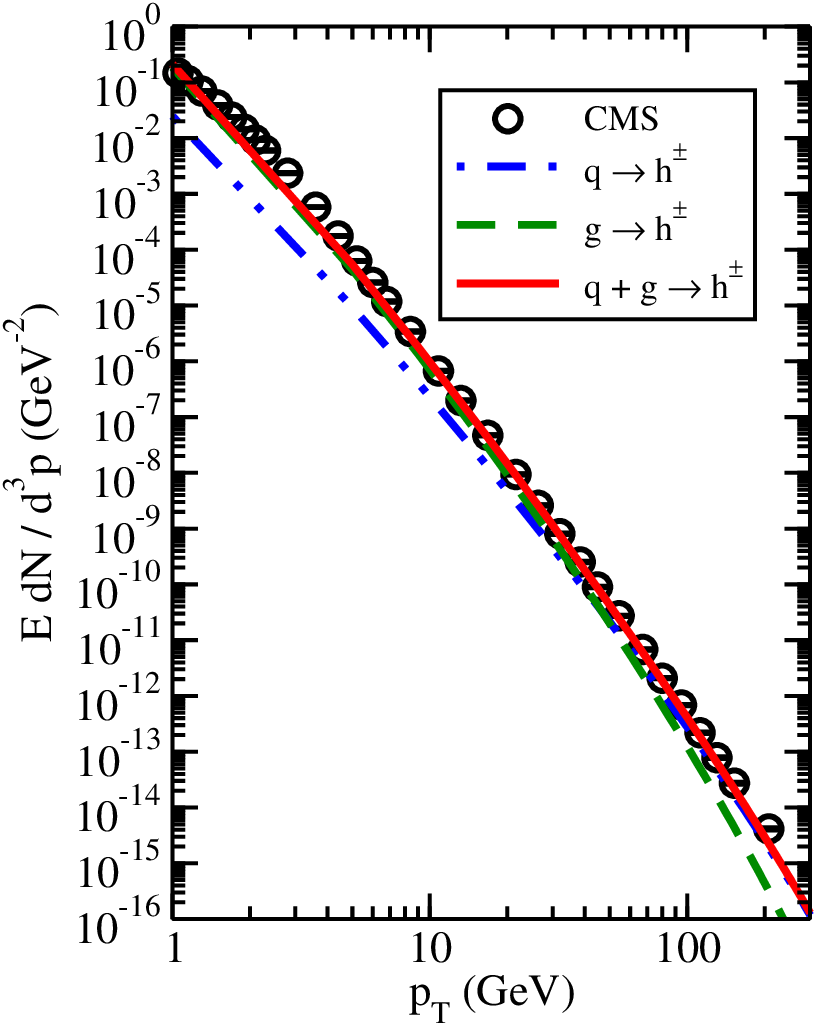}
\includegraphics[width=0.49\linewidth]{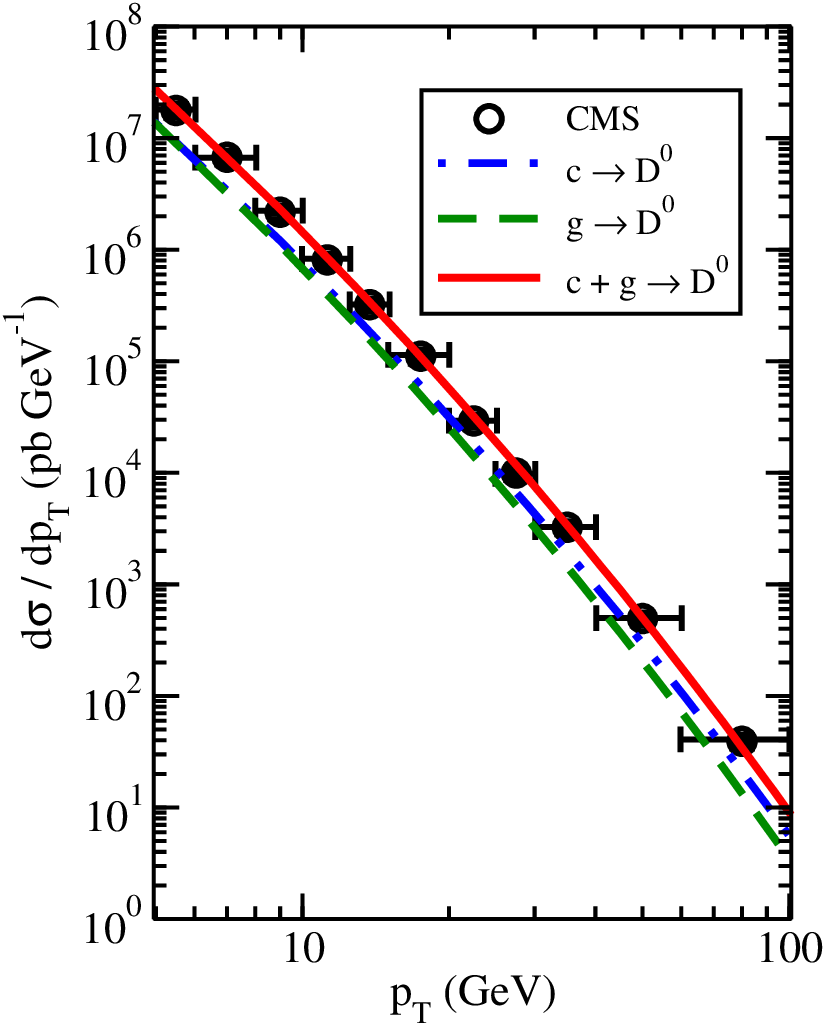}
\caption{Transverse momentum spectra for charged hadrons and $D$ mesons in p+p collisions at 5.02~TeV from NLO perturbative QCD calculation compared to the CMS data \cite{Khachatryan:2016odn, Sirunyan:2017xss}.
} \label{fig1}
\end{figure}

For inelastic radiative process, we use higher-twist energy loss formalism, in which the medium-induced gluon radiation spectrum takes the following form  \cite{Wang:2001ifa, Zhang:2003wk, Majumder:2009ge},
\begin{eqnarray}
\frac{dN_g^a}{dxdl_\perp^2 dt} = \frac{2 C_A \alpha_s P_a(x) l_\perp^4 \hat{q}_a}{\pi (l_\perp^2 + x^2M^2)^4} \sin^2\left(\frac{t-t_i}{2\tau_f}\right)
\end{eqnarray}
where $M$ is mass of the propagating parton, $x$ and $l_\perp$ are the momentum fraction and transverse momentum carried by the radiated gluon with respect to the parent parton, $\alpha_s$ is the strong coupling for the splitting vertex, $P_a(x)$ is the splitting function, $\hat{q}_a$ is the transport coefficient (the transverse momentum transfer squared per mean free path) due to elastic scatterings between the propagating parton and medium constituents, $t_i$ is the time of the last gluon radiation, and $\tau_f = 2Ex(1-x)/(l_\perp^2+x^2M^2)$ is the gluon formation time, with $E$ being the energy of the parent parton.
Here we take light partons to be massless, and for heavy quarks, we use $M_c = 1.27$~GeV and $M_b = 4.19$~GeV.
The above gluon radiation spectrum for the propagating parton $a$ is used to calculate the inelastic scattering rate $\Gamma_{\rm inel}^a =  \int dx dl_\perp^2 (dN_g^a/dxdl_\perp^2 dt) / (1+\delta_g^a)$ with the medium constituents, the average number of emitted gluons $\langle N_g^a \rangle = \Gamma_{\rm inel}^a \Delta t$, and the inelastic scattering probability $P_{\rm inel}^a = 1 - e^{-\langle N_g^a \rangle}$, in a given time step $\Delta t$.

In the LBT model, the total scattering probability $P_{\rm tot}^a = P_{\rm el}^a + P_{\rm inel}^a - P_{\rm el}^a P_{\rm inel}^a$ is splitted into two parts, the probability for pure elastic scatterings $P_{\rm el}^a (1-P_{\rm inel}^a)$ and the probability for inelastic scatterings with at least one gluon emission $P_{\rm inel}^a$.
These probabilities combined with the information about jet partons and medium profiles are used in our Monte-Carlo model to simulate the evolution of jet partons in the QGP medium.
More details on the LBT model can be found in Refs.~\cite{Cao:2017hhk, Cao:2016gvr, He:2015pra}.

{\it Numerical results } -- We first show, in Fig.~\ref{fig1}, the transverse momentum spectra for light charged hadrons and $D$ mesons in proton-proton collisions at 5.02~TeV based on the NLO perturbative QCD calculation \cite{Jager:2002xm, Aversa:1988vb}, compared to the CMS data \cite{Khachatryan:2016odn, Sirunyan:2017xss}.
The factorization scale and the renormalization scales are all taken as the jet parton $p_{\rm T}$ in the calculation.
One can see that the NLO perturbative QCD calculation can provide a very good description of both charged hadron and $D$ meson spectra (at relatively high $p_{\rm T}$). In the figure, we also show the relative contributions from quark and gluon fragmentations to charged hadron and $D$ meson productions. For charged hadrons, gluon contribution is more dominant at low $p_{\rm T}$, and quark contribution becomes more important at $p_{\rm T}>50$~GeV.
For $D$ mesons, charm quark fragmetation and gluon fragmentation contribute almost equally to the $D$ meson yield at low $p_{\rm T}$. Then with increasing $p_{\rm T}$ of $D$ mesons, the gluon contribution decreases, but it still renders around 40\% contribution to the $D$ meson yield at $p_{\rm T}=100$~GeV.
Note that the NLO perturbative QCD framework adopted here uses the zero-mass factorization scheme, thus is not valid for very small $p_{\rm T}$.

\begin{figure}[tb]
\includegraphics[width=0.96\linewidth]{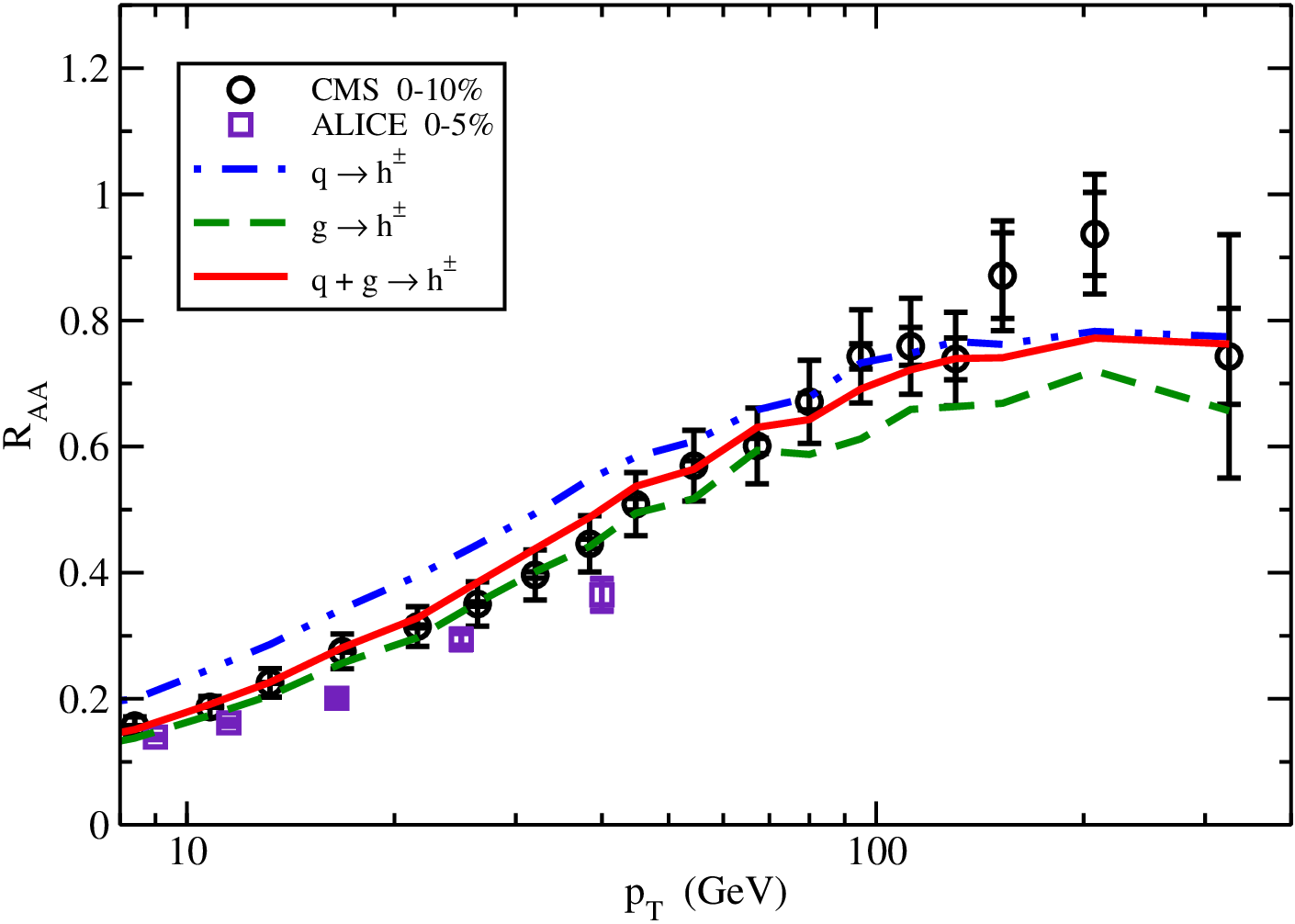}
\caption{Charged hadron $R_{\rm AA}$ as a function of $p_{\rm T}$ in central 0-10\% Pb+Pb collisions at 5.02A~TeV compared to the CMS 0-10\% and ALICE 0-5\% data \cite{Khachatryan:2016odn, Acharya:2018qsh}. Also shown are $R_{\rm AA}$'s for charged hadrons produced from light quarks and gluons, respectively.
} \label{fig2}
\end{figure}

In Fig.~\ref{fig2}, we show the nuclear modification factor $R_{\rm AA}$ as a function of $p_{\rm T}$ for charged hadrons in central 0-10\% Pb+Pb collisions at 5.02A~TeV at the LHC, compared to the CMS 0-10\% and ALICE 0-5\% data \cite{Khachatryan:2016odn, Acharya:2018qsh}.
In this study, the space-time evolution of the QGP fireball in 5.02A~TeV Pb+Pb collisions is obtained via a (3+1)-dimensional viscous hydrodynamics model CLVisc \cite{Pang:2012he, Pang:2018zzo} in which $\tau_0 = 0.6$~fm, $\eta/s = 0.08$ and $T_c = 165$~MeV are employed to describe the soft hadron spectra.
Note that the default version of the LBT model only considers leading-order $2\to 2$ elastic scattering processes, thus the distribution for the exchanged transverse momentum between jet partons and medium constituents typically has a hard power-law tail. To account for the possible contributions from multiple soft scatterings whose transverse momentum distribution is typically a Gaussian, we impose an effective momentum cutoff for transverse momentum exchange between jet and medium ($k_\perp < 10 T$).
Such setup reduces the energy loss of jet partons and requires larger value of strong coupling $\alpha_s$ as compared to Ref. \cite{Cao:2017hhk}.
In this work, the coupling for the interaction vertex connecting to thermal partons is taken as $\alpha_s=0.2$.
For the vertices connecting to jet partons, we take the running coupling as: $\alpha_s = 4\pi/[9\ln(2ET/\Lambda^2)]$, with $\Lambda = 0.2$~GeV.
More detailed study on the interplay between single hard and multiple soft scatterings and their influences on final-state observables will be explored in the future effort.
In the figure, we also show $R_{\rm AA}$'s for charged hadrons produced from light quarks and gluons, respectively. One can see that due to the color effect, quark-initiated hadrons exhibit less quenching effects than gluon-initiated hadrons. After combining both quark and gluon fragmentations to charged hadrons, our model gives a nice description of charged hadron $R_{\rm AA}$ over a wide range of transverse momenta ($p_{\rm T}=8$-$300$~GeV).

\begin{figure}[tb]
\includegraphics[width=0.96\linewidth]{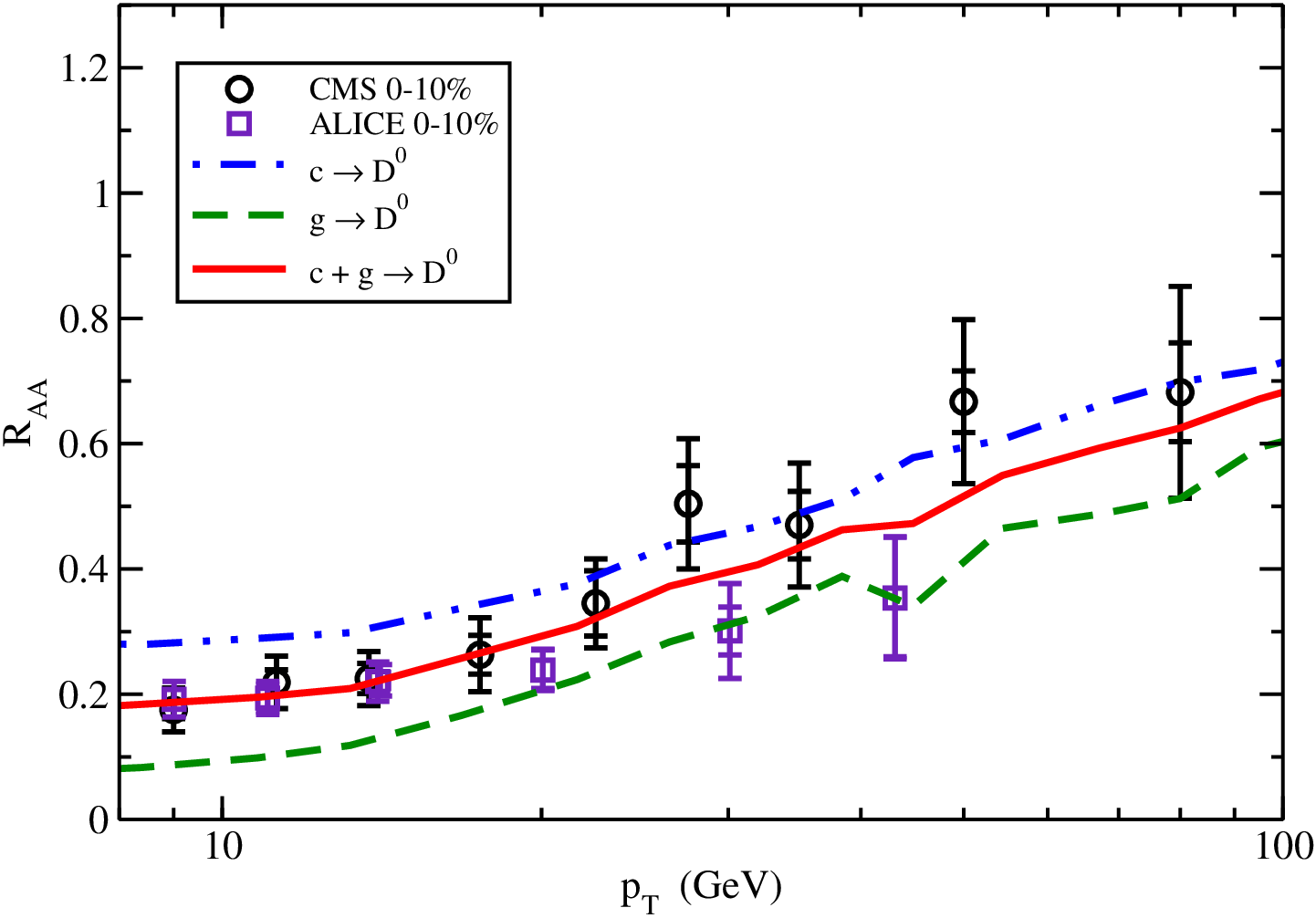}
\caption{$D$ meson $R_{\rm AA}$ as a function of $p_{\rm T}$ in central 0-10\% Pb+Pb collisions at 5.02A~TeV compared to the CMS and ALICE data \cite{Sirunyan:2017xss, Acharya:2018hre}. Also shown are $R_{\rm AA}$'s for $D$ mesons produced from charm quarks and gluons, respectively.
} \label{fig3}
\end{figure}

Figure~\ref{fig3} shows the nuclear modification factor $R_{\rm AA}$ as a function of $p_{\rm T}$ for $D$ mesons in central 0-10\% Pb+Pb collisions at 5.02A~TeV at the LHC, compared to the CMS and ALICE data \cite{Sirunyan:2017xss, Acharya:2018hre}. In the figure, we also show $R_{\rm AA}$'s for $D$ mesons produced from charm quarks and gluons, respectively. Similar to charged hadrons, we can see that $D$ mesons produced from charm quark fragmentation have less quenching than $D$ mesons from gluon fragmentation. Again, after combining both charm quark and gluon contributions to $D$ meson production, we obtain successful description of $D$ meson $R_{\rm AA}$ data from CMS for $p_{\rm T}=$~8-100~GeV.

\begin{figure}[tb]
\includegraphics[width=0.96\linewidth]{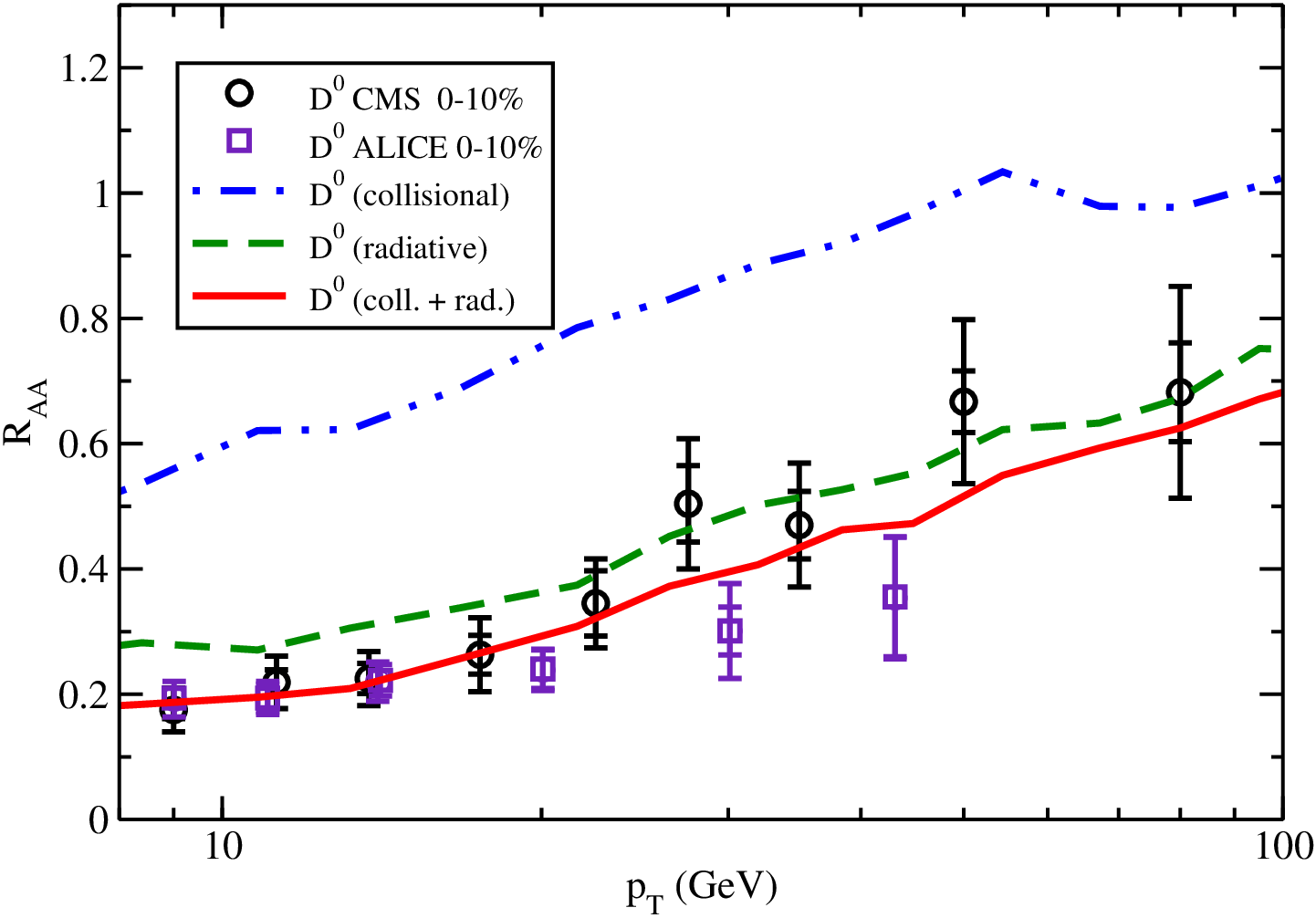}
\caption{Relative contributions from collisional and radiative energy loss components to $D$ meson $R_{\rm AA}$ as a function of $p_{\rm T}$ in 0-10\% Pb+Pb collisions at 5.02A~TeV.
} \label{fig4}
\end{figure}

In our study, both elastic scattering and inelastic radiative processes are included in the LBT simulation. The relative contributions from collisional and radiative energy loss components to the nuclear modifications of $D$ mesons are shown in Fig.~\ref{fig4} for central 0-10\% Pb+Pb collisions at 5.02A~TeV at the LHC. One can see that while radiative energy loss provides more dominant contributions to the nuclear modification factor $R_{\rm AA}$ in the $p_{\rm T}$ range explored here, collisional energy loss also gives sizable contributions to $R_{\rm AA}$ at not-very-high $p_{\rm T}$ regime and such contribution diminishes with increasing $p_{\rm T}$.

\begin{figure}[tb]
\includegraphics[width=0.96\linewidth]{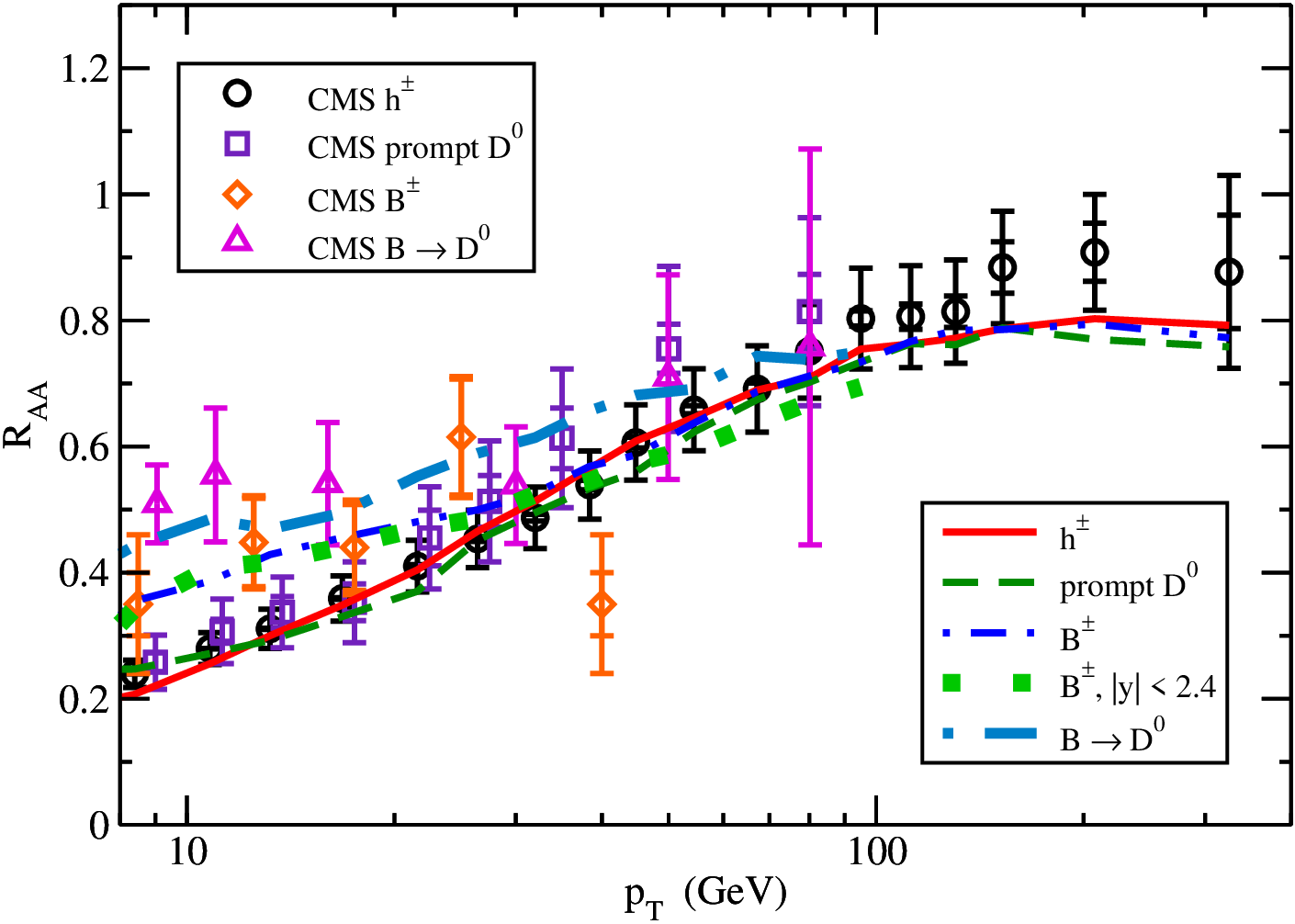}
 \caption{Nuclear modification factors for charged hadrons, prompt $D$ mesons, $B$ mesons and $B$-decayed $D$ mesons in 0-80\% Pb+Pb collisions at 5.02A~TeV compared to the CMS minimum bias data \cite{Khachatryan:2016odn, Sirunyan:2017xss, Sirunyan:2017oug, Sirunyan:2018ktu}.
} \label{fig5}
\end{figure}

The above results clearly show that our calculation can simultaneously describe both $D$ meson and light charged hadron $R_{\rm AA}$'s for central 0-10\% Pb+Pb collisions at 5.02A~TeV at the LHC.
Recently, CMS Collaborations have measured the nuclear modification factors for $B$ mesons up to 30-40~GeV and $B$-decayed $D$ mesons up to 80-100~GeV in Pb+Pb collisions at 5.02A~TeV \cite{Sirunyan:2017oug, Sirunyan:2018ktu}.
It is very interesting to see whether our model can describe $B$ meson and $B$-decayed $D$ meson suppressions as well since beauty quarks have much larger mass.
In Fig.~\ref{fig5}, we show the nuclear modification factor $R_{\rm AA}$ as a function of $p_{\rm T}$ for $B$ mesons and $B$-decayed $D$ mesons, together with $R_{\rm AA}$'s for charged hadrons and prompt $D$ mesons, for 0-80\% Pb+Pb collisions at 5.02A~TeV at the LHC. Also shown are the CMS minimum bias data \cite{Khachatryan:2016odn, Sirunyan:2017xss, Sirunyan:2017oug} for comparison. It is worth noting that we compute $R_{\rm AA}$ for 0-80\% as follows: $\langle R_{\rm AA} \rangle = \sum_{c} P^{(c)}  R_{\rm AA}^{(c)}$, where $P^{(c)} = N_{\rm bin}^{(c)} / \sum_{c} N_{\rm bin}^{(c)}$ is the probability of finding jet events in a given centrality bin. If one uses an average medium profile via averaging the hydrodynamics profiles or initial conditions over different centralities, much less jet quenching effects would be obtained for the minimum bias calculation.
Note that there are two $B$ meson $R_{\rm AA}$ curves in the figure: one with rapidity $|y|<1$ to compare with charged hadrons and $D$ mesons, another with $|y|<2.4$ to compare with the CMS data.
From the figure, one can see that our model can simultaneously describe the nuclear modifications of charged hadrons, prompt $D$ mesons, $B$ mesons and $B$-decayed $D$ mesons.
Below $p_{\rm T}=$~30-40~GeV, $B$ mesons exhibit less quenching than charged hadrons and $D$ mesons, while above 30-40~GeV, our model predicts similar quenching for $B$ mesons to charged hadrons and $D$ mesons.
We have verified that our conclusion on the flavor dependence of jet quenching is robust against various theoretical uncertainties, such as the factorization and renormalization scales, the lower temperature cutoff for jet-medium interaction, the cut for transverse momentum exchange between jet and medium in elastic collisions, etc.
Future LHC experiments should be able to test our result.

{\it Summary} -- With our state-of-the-art jet quenching model, we have presented a systematic and most complete study on the energy loss and nuclear modification of heavy and light flavor jets and hadrons in high-energy heavy-ion collisions. Our model combines a NLO perturbative QCD framework to calculate the productions of high $p_{\rm T}$ jet partons and hadrons, a linear Boltzmann transport model to simulate the evolution of heavy and light flavor jet partons in the QGP, and a realistic hydrodynamic model to describe the space-time evolution of the QGP fireball.
It not only includes quark and gluon contributions to light and heavy flavor hadron productions, but also incorporates both elastic and inelastic interactions between jet partons and the medium constituents.
By incorporating all important ingredients in our perturbative QCD based jet quenching model, we obtain the first satisfactory description of the experimental data for the nuclear modification factors of charged hadrons, prompt $D$ mesons, $B$ mesons and $B$-decayed $D$ mesons over the widest range of transverse momenta ($p_{\rm T}=$~8-300~GeV) in literature.
This provides a natural solution to the flavor hierarchy puzzle of jet quenching, and constitutes a significant step forward for the precision study of jet-medium interaction in relativistic heavy-ion collisions.
Our study also demonstrates that perturbative QCD calculation is sufficient for studying the color, mass and energy dependence of parton energy loss and jet quenching.
With a solid understanding on how jet-medium interaction depends on jet properties (color, mass and energy), we are now in a position to really use jets to quantitatively probe the QGP properties.
Based on our jet quenching model calculations, we predict that at transverse momenta $p_{\rm T}>$~30-40~GeV, $B$ mesons will also exhibit similar suppression effects to charged hadrons and $D$ mesons, which can be tested by future high luminosity precision measurements.


{\it Acknowledgments} -- We thank Yayun He, Zhongbo Kang, Tan Luo and Xin-Nian Wang for very helpful discussions. This work is supported in part by Natural Science Foundation of China (NSFC) under Grants No. 11775095, No. 11890711 and No. 11375072, and by the China Scholarship Council (CSC) under Grant No. 201906775042. S. C. is supported by U.S. Department of Energy under Contract No. DE-SC0013460. H. X. is supported by NSFC under grant No. 11435004.

\bibliographystyle{plain}
\bibliographystyle{h-physrev5}
\bibliography{refs_GYQ}
\end{document}